\documentclass[12pt,leqno]{article}
\usepackage{amsmath,amsfonts,amssymb,amsthm}
\usepackage{graphicx}

\newtheorem{teo}{Theorem.}

\date{Proceeding of GDIS 2010, Belgrade}

\begin{document}
\title{Three and four-body systems in one dimension: \\ integrability, superintegrability and discrete symmetries}
\author{Claudia Chanu \\ \\
Dipartimento di Matematica e Applicazioni,\\ Universit\`a di Milano Bicocca.  Milano, via Cozzi 53, Italia. 
\footnote{Research partially supported by the European program  ``Dote ricercatori'' (F.S.E. and Regione Lombardia)}
\\
\\
 Luca Degiovanni, \, Giovanni Rastelli.\\ \\ Formerly at Dipartimento di Matematica, \\ Universit\'a di Torino. \\ \\e-mail: claudia.chanu@unimib.it \\ luca.degiovanni@gmail.com \\ giorast.giorast@alice.it }

\maketitle

\abstract{Families of three-body Hamiltonian systems in one dimension have been recently proved to be maximally superintegrable by interpreting them as one-body systems in the three-dimensional Euclidean space, examples are the Calogero, Wolfes and Tramblay Turbiner Winternitz systems. For some of these systems, we show in a new way how the superintegrability   is associated with their dihedral symmetry in the three-dimensional space,    the order of the dihedral symmetries being associated with the degree of the polynomial in the momenta first integrals. As a  generalization, we introduce the analysis of  integrability and superintegrability of four-body systems in one dimension by interpreting them  as one-body systems  with the symmetries of the Platonic polyhedra in the four-dimensional Euclidean space. The paper is intended as a short review of recent results in the sector, emphasizing the relevance of discrete symmetries for the superintegrability of the systems considered.
}

\subsection*{Introduction}
In recent years several new classical and quantum superintegrable systems have been discovered, with additional polynomial first integrals of high degree. In many cases, these systems can describe the dynamics of point particles in one dimension with reciprocal interaction, as for instance Calogero-Moser or Wolfes systems.
A common feature of these systems is the existence of dihedral or polyhedral symmetry groups, in some cases related with the invariance of the system under permutations of the particles, when interpreted as dynamics of points on a line.
We shortly review some of the most recent results on this topic, pointing out the existing dihedral or polyhedral symmetries and how they can be related with additional first integrals. In particular we propose a class of quasi-maximally superintegrable systems in dimension four built from some of the Evans systems \cite{E} and we show how they can be considered as dynamics of four points on a line. As future developments we sketch an approach for building superintegrable four-dimensional systems with the symmetries of the Platonic polyhedra.

\subsection *{A class of superintegrable three-body systems on the line}
Natural Hamiltonian three-body (mass points) systems on a line, with positions $x^i$ and momenta $p_i$ can be interpreted as one-point systems in the three-dimensional Euclidean space with Cartesian coordinates and the standard metric, if the masses are all unitary \cite{CDR1}, (see \cite{BKM1} for a slightly different approach). The case of different positive masses is always reducible to the former \cite{BCR}. Let $X_i=x^i-x^{i+1}$, $\mathbf{\omega}=(1,1,1)$ and let $(r, \psi,z)$ be cylindrical coordinates with axis $\Omega$ parallel to $\mathbf{\omega}$ and passing through the origin. The natural Hamiltonians with scalar potentials of the form
\begin{equation}\label{E0}
V=\sum_i {\frac 1{X_i^2}F_i\left(\frac {X_{i+1}}{X_i},\frac {X_{i+2}}{X_i}\right)} =\frac {F(\psi)}{r^2},
\end{equation}
admit the four independent quadratic first integrals
$$
H=\frac 12\left(p_r^2+\frac 1{r^2}p_\psi^2+p_z^2\right)+\frac {F(\psi)}{r^2},\ \ \
H_1=\frac 12p_\psi^2+F(\psi), \ \ \ H_2=\frac 12 p_z^2,
$$
$$
H_3=\frac 12\left(rp_z-zp_r\right)^2+\frac {z^2}{r^2}H_1.
$$
The first three of the first integrals allow integration of the Hamilton-Jacobi equation by separation of variables and determine the radial motion common to any choice of $F$.
Examples are Calogero and Wolfes potentials
$$
V_C=\sum_{i=1}^3\frac {k}{X_i^2}=\frac k{\left( r\, \sin 3\psi \right)^2},\ \ 
V_W=\sum_{i=1}^3\frac {h}{\left(X_i-X_{i+1}\right)^2}= \frac h{\left( r\,\cos 3\psi \right)^2},
$$
with $(h,k \in R)$. Remarkably, a rotation of angle $\alpha=\frac \pi 6$ around the axis $\Omega$ of the cylindrical coordinates i.e., a phase shift $\psi\longrightarrow \psi+\alpha$, makes the separated equations for $V_C$, a two-body interaction,  and $V_W$, a three-body interaction, to coincide. Therefore, by reading the potentials as functions of $X_i$, the corresponding interactions among points on the line can be considered equivalent \cite{CDR1}. Potentials as $V_C$ and $V_W$ admit a further independent cubic first integral making them maximally superintegrable. In general, for every positive integer $n$, the Hamiltonian 

\begin{equation}\label{E00}
H_S=\frac 12 \left( p_r^2+\frac 1{r^2}p_\psi^2\right)+\frac F{r^2} \quad \text{with} \quad F=\frac k{\sin ^2 (n\psi+\psi_0)},
\end{equation}
i.e., $H_S=H-H_2$, admits the fifth constant of the motion 
\begin{equation}\label{E00bis}
L_n=\left[ p_r+\frac 1{n \, r}\left( p_\psi\frac {\partial}{\partial \psi}-\frac  {d F}{d\psi}\frac {\partial}{\partial p_\psi}\right) \right]^n \cos (n \psi+\psi_0)
\end{equation}
which is a polynomial  of degree $n$ in $(p_r,p_\psi)$ and is obviously a first integral of $H$ too \cite{CDR3}. The function $L_{3}$ coincides with the well known cubic first integral of the Calogero system. The procedure of above has been generalized in \cite{CDR4} to Hamiltonians of the form
\begin{equation}\label{Ext}
H=\frac 12 p_u^2+\alpha(u)L,
\end{equation}
where $L$ is an $n$-dimensional natural Hamiltonian on a (pseudo)- Riemannian manifold $Q^n$ and $\alpha$ is a function of $u$ only. The main result can be summarized as follows

\teo\label{Teo}
Let  $Q^n$ be a $n$-dimensional (pseudo-)Riemannian manifold with metric tensor $\mathbf g$. 
A natural Hamiltonian $L=\frac{1}{2} g^{ij}p_ip_j+V(q^i)$ on $M=T^*Q$ with canonical coordinates $(p_i,q^i)$ admits an extension $H$ in the form (\ref{Ext}) 
with an additional first integral $U^m(G)$ with $U=p_u+ \gamma(u) X_L$, where $X_L$ is the Hamiltonian flow of $L$ and $G(q^i, a_i)$ depends on $n+1$ free parameters $(a_i)$, if and only if the following conditions hold:
\begin{enumerate}
\item $Q$ is a (pseudo)-Riemannian manifold with constant curvature $K$; 
\item
the functions $G$ and $V$ satisfy
\begin{equation}\label{HessTeo} 
 \mathbf{H}(G)\,=\,-K\, \mathbf{g}G,
\end{equation}
\begin{equation}
\label{VTeo} \nabla V \cdot \nabla G-2KVG=0,
\end{equation}
where $\mathbf H(G)_{ij}=\nabla_i\nabla_jG$ is the Hessian tensor of $G$.
\item for $K=0$ the extension is  
$H=\tfrac 12 p_u^2-\kappa L$ ($\kappa \in \mathbb R$)

for $K\neq 0$ the extension is of the form $$H=\frac 12 p_u^2+ \frac{K}{S_\kappa^2(cu+u_0)} L$$
where $\kappa$, $u_0$ are arbitrary constants, $c=K/m$ and 
$$
S_\kappa(x)=\left\{\begin{array}{ll}
\frac{\sin\sqrt{\kappa}x}{\sqrt{\kappa}} & \kappa>0 \\
x & \kappa=0 \\
\frac{\sinh\sqrt{|\kappa|}x}{\sqrt{|k|}} & \kappa<0
\end{array}\right.
$$
\end{enumerate}

\rm

Superintegrable systems with higher order polynomial first integrals in constant curvature manifolds are considered also in \cite{MPY}, where  the constant curvature condition is assumed, and in \cite{KMK1}, where the remarkable property that all superintegrable systems in two-dimensional (pseudo-)Riemannian manifolds are St\"ackel-equivalent to  superintegrable systems in constant curvature manifolds is taken in account. In Theorem \ref{Teo} the constant curvature condition is derived as integrability condition for (\ref{HessTeo}). 

Recently \cite{TTW1, TTW2}, the family of systems (\ref{E00}) has been generalized into the so-called Tramblay-Turbiner-Winternitz (TTW)  systems with
$$
V_{TTW}=\frac 1{r^2}\left[k_1+\frac {k_2}{(\cos h\psi)^2}+\frac {k_3}{(\sin h\psi)^2}\right].
$$
where $h=\dfrac pq$ is a rational coefficient. They have been proved superintegrable with polynomial first integrals for all $k_i\in \mathbb R$ and $h\in \mathbb Q$, \cite{KMK1,KMK2, KMK3} and \cite{MPY, BKM2}; in all these articles, superintegrability is proved by using the separability of the TTW system to build explicitly an extra first integral. This connection between separable (St\"ackel) systems and superintegrability is deeply analyzed from an algebraic point of view in \cite{T}. In \cite{MPT} new bi-Hamiltonian superintegrable two-dimensional systems with higher-order first integrals are obtained, both separable and not, by generalizing methods exposed in \cite{MPY} and \cite{T}.

\subsection*{Dihedral symmetries} 

Any potential of the form $V=\frac k{[r\, \sin n\psi]^2}$ in $\mathbb R^3$
is invariant under the dihedral symmetry group of the prism having as basis the regular polygon with $2n$ sides; for $n=3$ the hexagon (Calogero and Wolfes case), for $n=4$ the octagon etc. The potentials of the TTW systems with rational coefficients $h=\frac pq$, represented in polar coordinates on the Euclidean plane, 
are not single-valued for $q>2$ but still show, at least formally, dihedral symmetry of order $p$ or $2p$ respectively.
Indeed, for $k_2\neq k_3$, they have period $q\pi$, if $q$ is even, or $2q\pi$ otherwise; for $k_2=k_3$, dihedral symmetries and periods are the same as above with $2h$ instead of $h$: by substituting 
$$
\cos^2 (h\psi) =(1+\cos (2h\psi))/2, \quad \sin^2 (h\psi)=(1-\cos (2h\psi))/2, 
$$
in $V_{TTW}$, one obtains 
$$
V_{TTW}=\frac 1{r^2}\left[k_1+2\frac {(k_2+k_3)+(k_3-k_2)\cos 2h\psi}{\sin ^2 2h\psi}\right].
$$
This function has the symmetries and the period of $\cos 2h\psi$, the same as $\sin ^2 h\psi$, unless $k_2=k_3$ when the system
reduces to a TTW system with  parameter $2h$. 
It is remarkable that in literature no use is made of the discrete symmetries of TTW  for proving the integrability or superintegrability of the system through the reduction of the whole system to one of the sectors equivalent up to the symmetries, sectors that are dynamically separated because of the singularities of the potential. For systems of point particles on the line, such as the Calogero-Wolfes system, the dihedral symmetry is a natural consequence of the invariance of the Hamiltonian under permutations of the particles. The representation of the three body system as one-body system in the  three-dimensional space makes the symmetry immediately recognizable.
Potentials of the form (\ref{E0}) with $F=\frac k{\sin ^2 n\psi}$ and $n>3$, for example, still represent the dynamics of three points on a line, but the connection of their higher order of dihedral symmetry with permutations of the particles has been not  investigated yet.

\section*{Systems of $n$ points on a manifold} Any homogeneous function of degree $-2$ in $X_i=x^i-x^{i+1}$, $i=1,\ldots,n-1$, can be written in the form
\begin{equation}\label{E1}
V=\frac 1{X_1^2}F\left(\frac {X_2}{X_1},\ldots,\frac{X_{n-1}}{X_1}\right)=\frac 1{r_{n-1}^2}\Phi(\psi_1,\ldots, \psi_{n-2}),
\end{equation}
and viceversa, where $\tan \psi_i=\frac {X_{i+1}}{X_1}$ and $(r_{n-1},\psi_1,\ldots,\psi_{n-2},u)$ are spherical- cylindrical coordinates  in $\mathbb E^n$. 
The coordinate change from $(x^i)$ to the spherical-cylindrical coordinates can be obtained by introducing, for example, coordinates $(z^j)$  in $\mathbb R^n$ defined by
\begin{eqnarray*}
z^j&=&\frac 1{\sqrt{j(j+1)}}(x^1+x^2+\ldots +x^j-jx^{j+1}) \quad j=1\ldots n-1\\
z^n&=&\frac 1{\sqrt{n}}(x^1+\ldots+x^n),
\end{eqnarray*}
which are Cartesian coordinates equioriented with $(x^i)$, and then to the $(r_{n-1}, \psi_i,u)$ where $u=z^n$ and the other are standard $n-1$-dimensional spherical coordinates. The $n$-body system in one dimension becomes in this way an one-point system in the $n$-dimensional Euclidean space. The literature about the superintegrability of systems of this kind is very rich, the procedure of above allows to find immediately  new integrable or superintegrable systems of points on the line. As an example, let us consider Evans systems. All superintegrable potentials with at least four quadratic first integrals in $\mathbb E^3$ are listed in \cite{E}; those in the form (\ref{E1}) are, up to isometries,
\begin{eqnarray*}
V_1&=&\frac {F(\psi_1)}{(r_3\sin \psi_2)^2}, \\ 
V_2&=&\frac k{(r_3 \cos \psi_2)^2}+\frac {F(\psi_1)}{(r_3 \sin \psi_2)^2},\\ 
V_3&=&\frac {k\cos \psi_2+F(\psi_1)}{(r_3 \sin \psi_2)^2},
\end{eqnarray*}
with  $r_3^2=x^2+y^2+z^2$. Some other particular cases are possible, but no undetermined functions are then included; for example
$$
V_4=\frac k{r_3^2}\left[k+\frac {k_1}{(\cos\psi_2)^2}+\frac 1{(\sin\psi_2)^2}\left(\frac {k_2}{(\cos \psi_1)^2}+\frac {k_3}{(\sin \psi_1)^2}\right)\right].
$$
All the potentials of above, trivially extended in dimension four by adding $\frac 12p_u^2$ to the three dimensional Hamiltonian $H$, admit six second degree independent first integrals in $\mathbb E^4$: the new Hamiltonian $\frac 12p_u^2+H$, 
\begin{equation} \label{H1}
H_1=\frac 12\left(p_{\psi_2}^2+\frac 1{(\sin \psi_2)^2}p_{\psi_1}^2\right)+r_3^2 V_i,
\end{equation}
which is an Hamiltonian on $\mathbb S^2$, the two other  first integrals of $H$ listed in \cite{E} and 
$$
H_5=p_u^2,\ \ H_6=\frac 12(up_r-r_3 p_u) ^2+\frac {u^2}{r_3^2}H_1.
$$
Due to the fact that these systems are in the form (\ref{E1}), it is always possible to write the functions of above in term of distances between four points on a line; again, phase shifts in $\psi_i$ determine equivalence classes of interactions on the line. Then, it is determined a class of superintegrable four-body systems on the line with $2n-2$ quadratic first integrals \cite{CDR2}.

Natural systems on $\mathbb E^n$ with potential of the form (\ref{E1}) can be interpreted as natural  systems of $n$ points on a line. Phase shifts $\psi_i\longrightarrow\psi_i+\alpha_i$, $0\leq \alpha_i< \pi$, determine equivalence classes of interactions on the line as seen in dimension three \cite{CDR1, CDR3}, \cite{BKM1}.
The same procedure can be extended to systems of $n$ points on a $m$-dimensional manifold. For example, the Hamiltonian of $n$ points of unitary masses in the Euclidean plane, denoted in Cartesian coordinates by $(x^1_a,x^2_a)$, $a=1,\ldots, n$, and moving under the force of  potential $V(x^i_a-x^i_b)$ can be considered as the Hamiltonian of one point in the $2n$ dimensional Euclidean space with Cartesian coordinates $q^{2a-1}=x_a^1,\, q^{2a}=x_a^2$, invariant with respect to the vectors $\omega_1=(1,0,1,\ldots,1,0)$ and $\omega_2=(0,1,0, \ldots,0,1)$ representing the conservation of the center of mass momentum.  Suitable spherical coordinates in the Euclidean space $\mathbb R^{2n-2}$ orthogonal to $\omega_1$ and $\omega_2$ can be introduced etc. The integrability or superintegrability of the system of $n$ points in the $m$-dimensional manifold can then be studied by considering the equivalent one-point system in the $nm$-dimensional manifold. Of course, this approach is not new and can be dated back to Lagrange itself.
In the case of four points on a line, it is remarkable the result of \cite{HNY, HKLN}, where the 4-body Calogero system is interpreted as a Cuboctahedric Higgs oscillator on $\mathbb S^2$ with Hamiltonian
\begin{eqnarray*}
H&=&\frac 12 \left( p_\theta^2+\frac{p_\phi^2}{\sin^2 \theta}\right)
+\frac{4g}{\sin^2 \theta}\left[\frac 1{1+\cos 4\phi}+ \right. \\
 &+&\left. \frac{k-6}{k-8+8/k-k\cos 4\phi} + \frac{4(k-16+16/k)}{(k-8+8/k-k\cos 4\phi)^2}\right],
\end{eqnarray*}
where $k=\tan^2 \theta$. This is the dynamics of one point under the potential generated by centers of force placed at the vertices of a cub-octahedron, an Archimedean polyhedron  with the symmetry of the octahedron. The connection between Higgs oscillators and (\ref{E00}) becomes evident due to the following elementary relations
$$
V_\text {Higgs}=a \tan^2\psi = a\left( \frac 1 {\cos^2 \psi}-1 \right), \quad a\in \mathbb R.
$$
The system is superintegrable as consequence of the superintegrability of the corresponding Calogero system. This fact, together with the dihedral symmetry of the three-body Calogero system, suggests strong links between the superintegrability of the $n$-body Calogero system \cite{W} and the existence of discrete symmetry groups possibly in connection with the invariance of the Calogero $n$-body Hamiltonian under permutations of the particles \cite{HNY}. For the interpretation of (\ref{E00}) as dynamics of one point on a sphere under the potential of equally spaced Hooke centers placed on a great circle, see \cite{BKM1,BKM2}.

\subsection*{Future directions: Platonic symmetries} As seen above, there is some strong connection between discrete symmetries and higher order polynomial first integrals of Hamiltonian systems. In \cite{R} is roughly analyzed the possibility of finding similar connections for Hamiltonians with Platonic symmetries as follows.  The Platonic symmetries are the symmetry groups of rotations  leaving invariant the platonic polyhedra that, apart the Tetrahedron, can be arranged in dual couples (the vertices of the one are the middle points of the faces of the other) sharing the same symmetry group:

Tetrahedron : $T_{12}$ or $A(4)$ 

Exahedron (Cube), Octahedron : $O_{24}$ or $S(4)$ 

Dodecahedron, Icosahedron : $I_{60}$ or $A(5)$ 

\begin{figure}
\begin{minipage}[b]{ 390pt} 
\begin{center}
\includegraphics*[width=   180pt, height=   180pt]{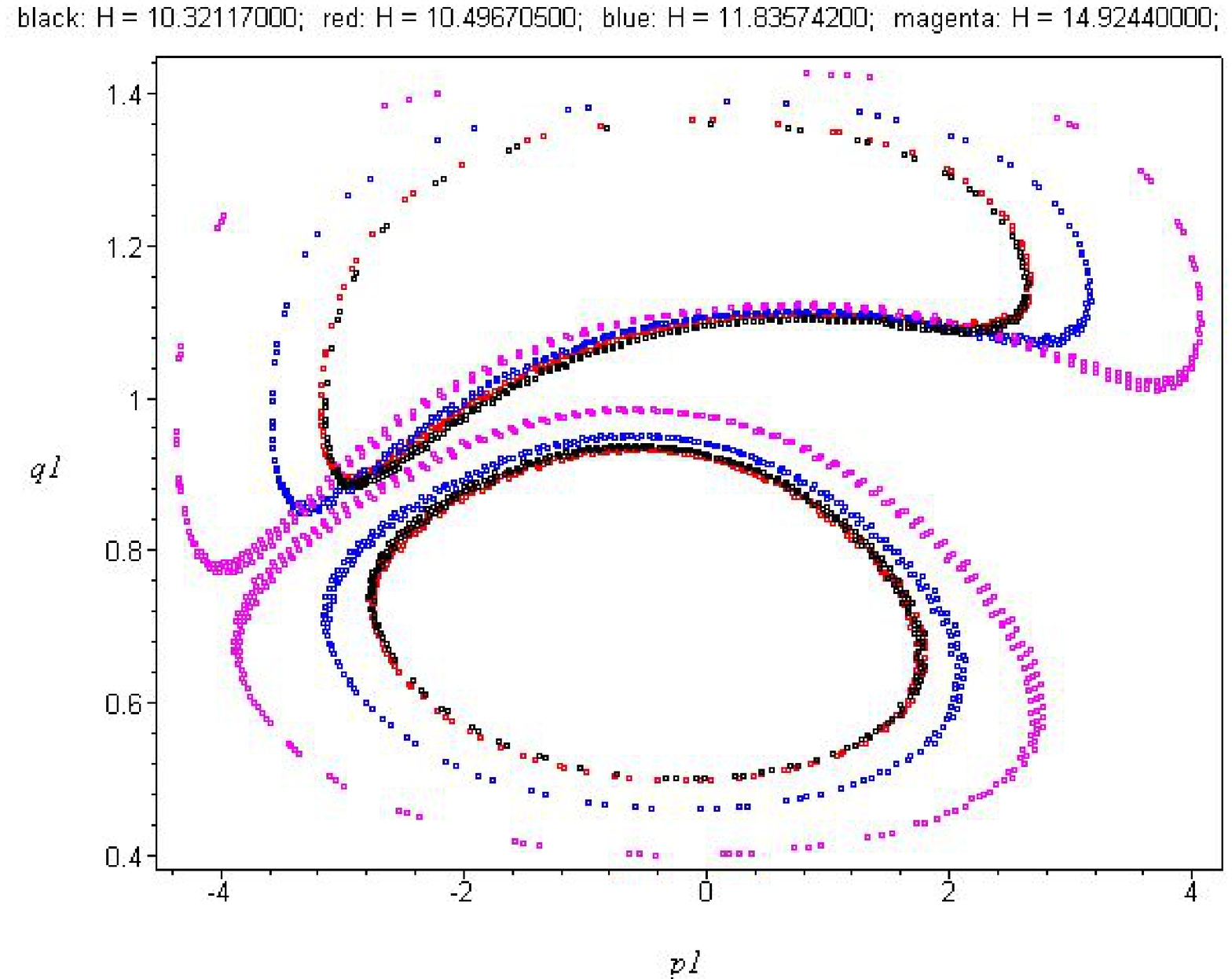}\includegraphics*[width=   180pt, height=   180pt]{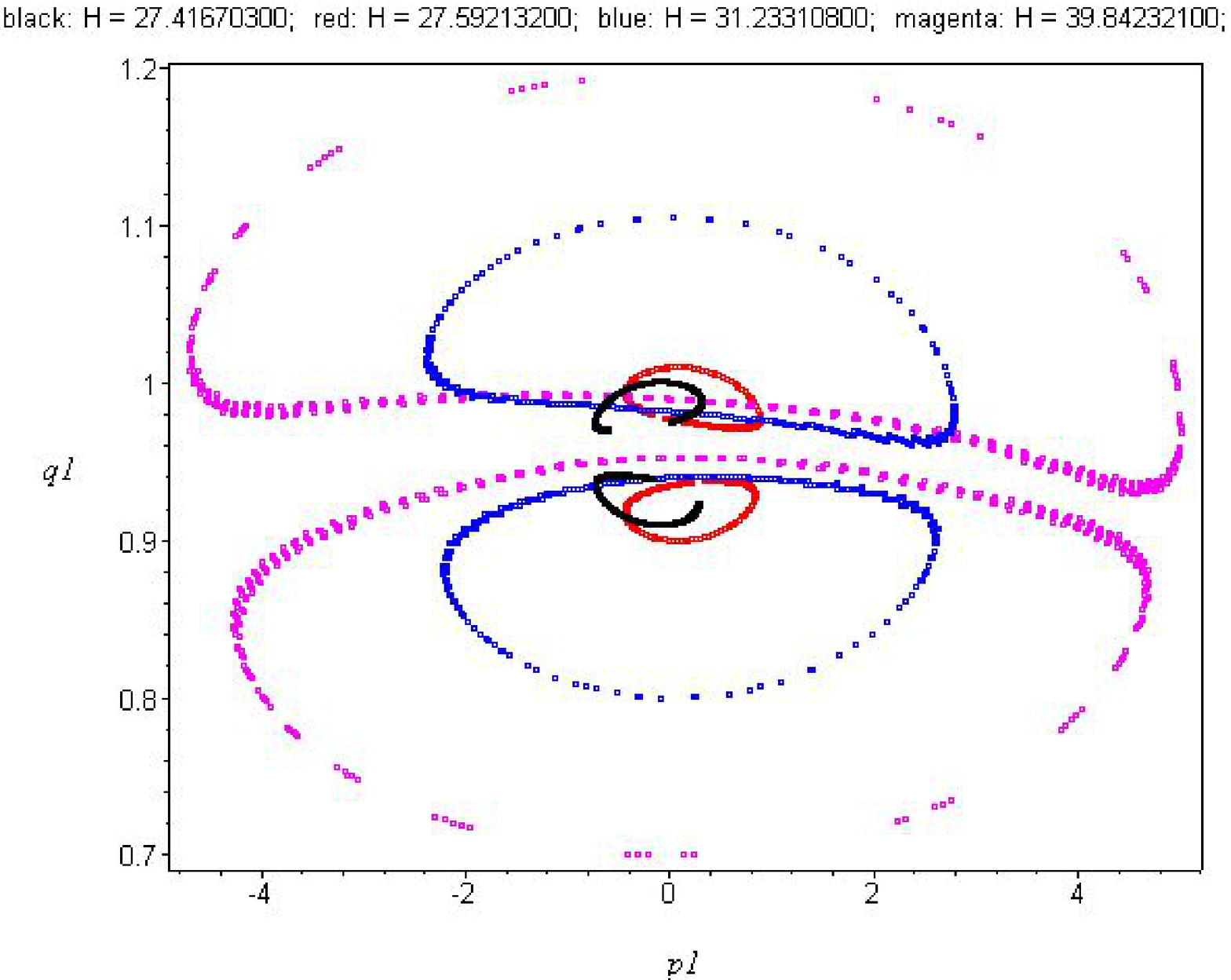}
\includegraphics*[width=   180pt, height=   180pt]{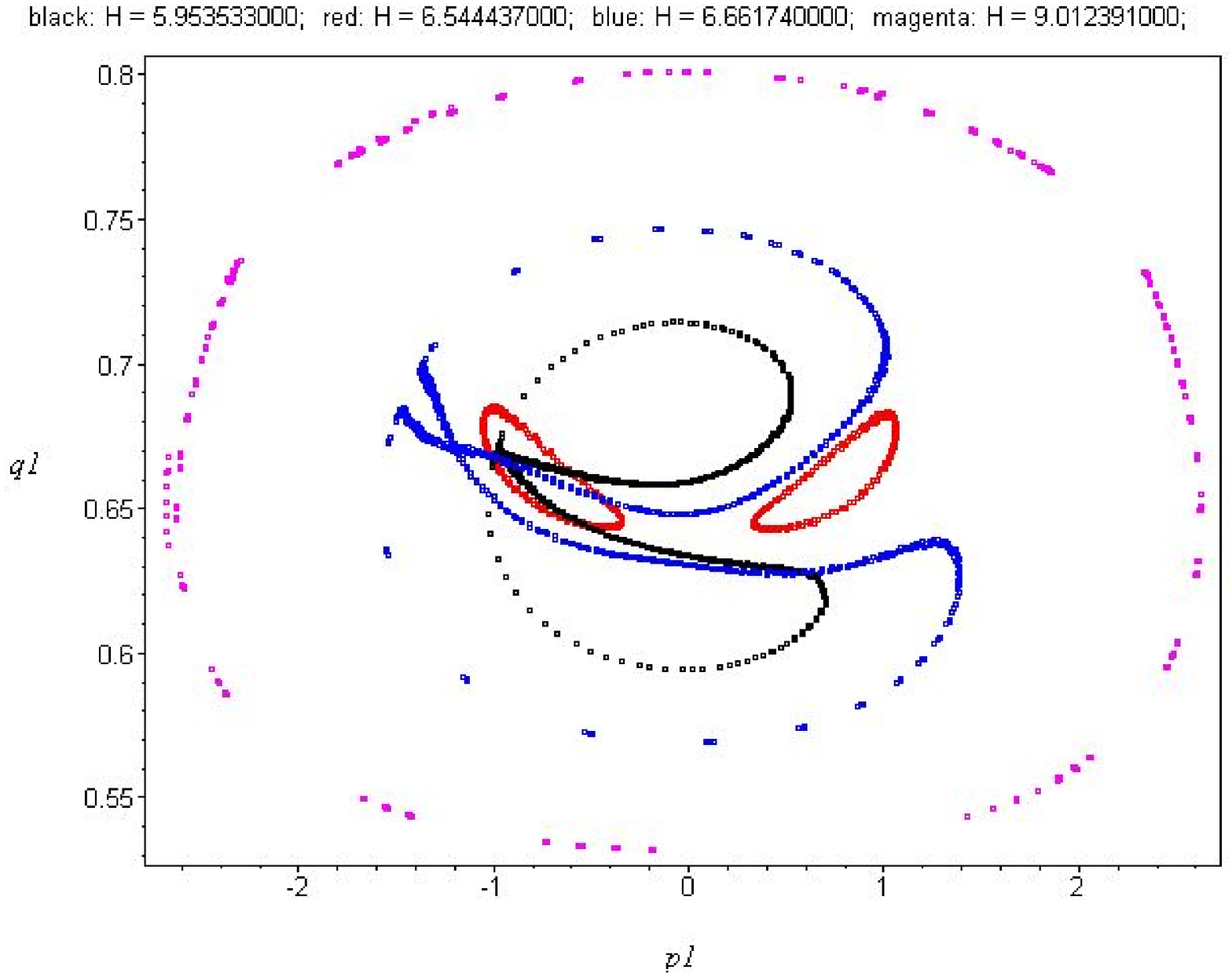}
\end{center}
\caption{{\it Here are shown  examples of Poincar\'e sections, with $q_1=\theta$, $p_1=p_\theta$ the momentum conjugate to $\theta$, of integral curves of  natural systems on the sphere with potential, from top left, $V_1$, $V_2$ and $V_3$ with four distinct initial condition sets, distinct by their colors. After some minimal level of accuracy in the numerical integration of the integral curves, their intersection points with a plane $(p_1,q^1$) in the phase space describe closed curves, a behaviour compatible with complete integrability.}\label{Ft1}}
\end{minipage}
\end{figure}

All invariant polynomials in $\mathbb E^3$ for each one of the symmetry groups are generated by a finite base of homogeneous polynomials in $(x,y,z)$: $(U_1,U_2,U_3)$ (by the celebrated Hilbert theorem).
In spherical coordinates these polynomials become
$$
U_i=r_3^{k_i}f_i(\theta, \psi),
$$
where $f_i$ are functions on $\mathbb S^2$ invariant for the corresponding platonic symmetry group. Among the simplest for each symmetry group, we have with tetrahedral symmetry:
$$
f_1=\sin^2\theta\cos\theta\cos\psi\sin\psi,
$$ 
with cubic-octahedral symmetry:
$$
f_2=f_1^2
$$ 
and with dodecahedral-icosahedral symmetry:
\begin{eqnarray*}
f_3=&-\cos\theta [ \cos^5\theta-5\sin^2\theta\cos^3\theta+5\sin^4\theta\cos\theta+\cr
&\sin^5\theta( 32\cos\psi\sin^4\psi-24\cos\psi\sin^2\psi+2\cos\psi)].
\end{eqnarray*}
If we consider the natural Hamiltonian in $\mathbb E^3$ with potentials of the form
$$
V_i=\frac 1{r_3^2\, f_i},
$$
we obtain  the  analogous in $\mathbb E^3$ with polyhedral symmetry of the potentials  in $\mathbb E^2$ with dihedral symmetry previously considered. On $\mathbb S^2$ the induced Hamiltonians ${\tilde H}_i$ are of the form (\ref{H1}) and, if integrable there, the whole three-dimensional systems is then superintegrable, as seen for the Evans systems above, and extensible to a superintegrable system in four-dimension corresponding to a four-body system in one dimension. There are numerical evidences that the Hamiltonians ${\tilde H}_i$  of above are integrable on $\mathbb S^2$, as shown by their Poincar\'e sections (Fig. \ref{Ft1}).

\begin{figure}
\begin{minipage}[b]{ 390pt} 
\begin{center}
\includegraphics*[width=   180pt, height=   180pt]{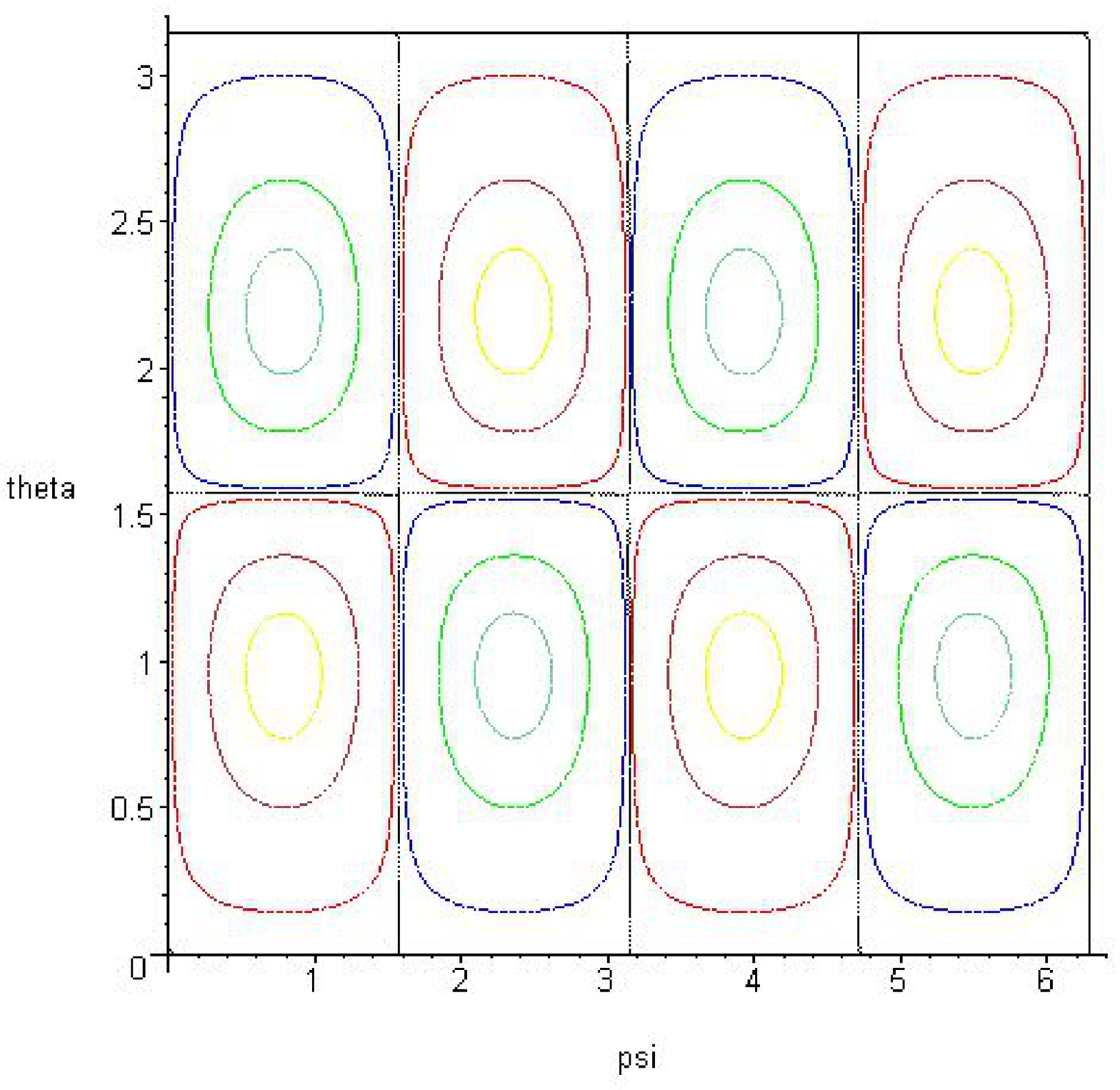}\includegraphics*[width=   180pt, height=   180pt]{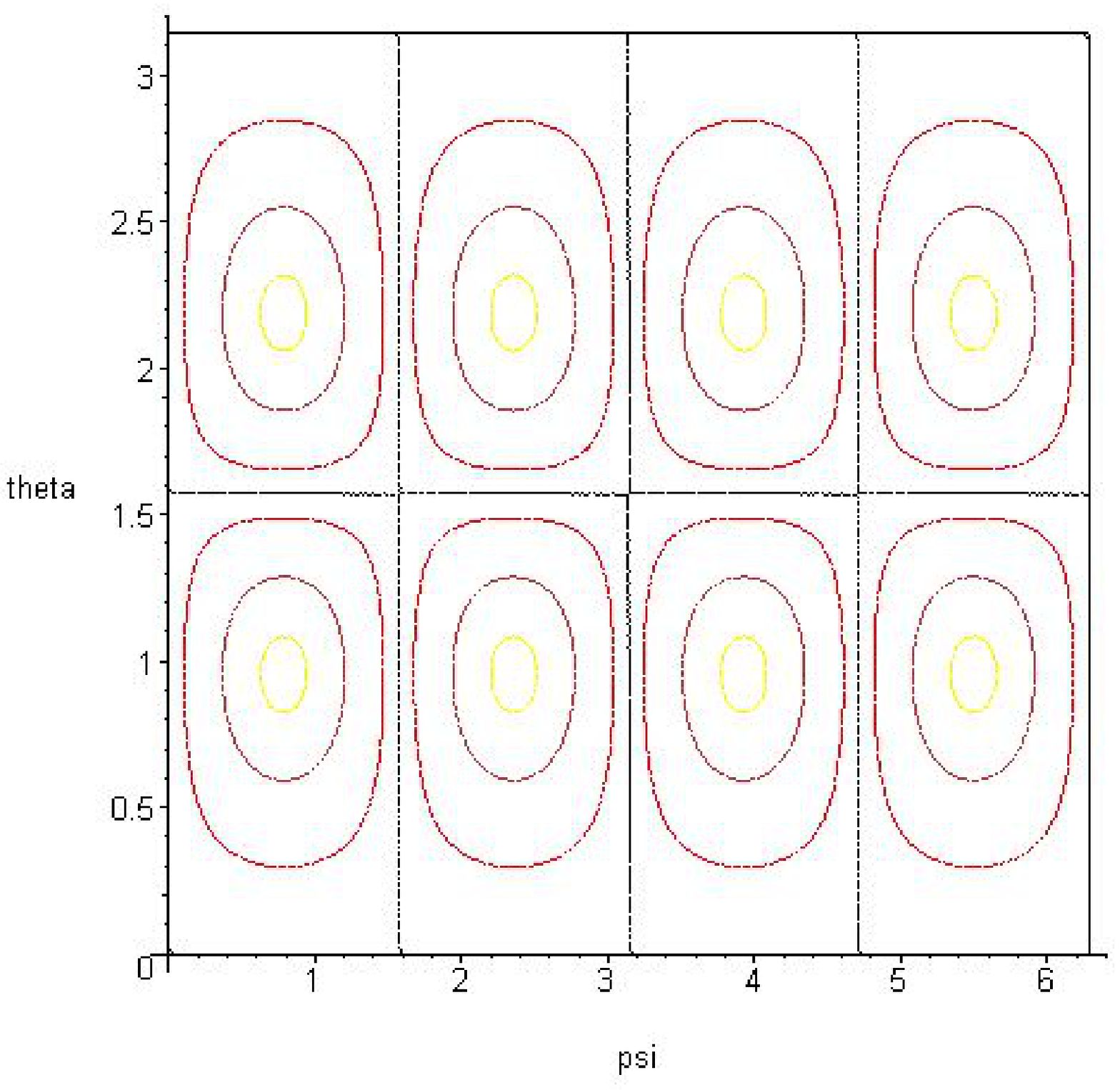}
\includegraphics*[width=   180pt, height=   180pt]{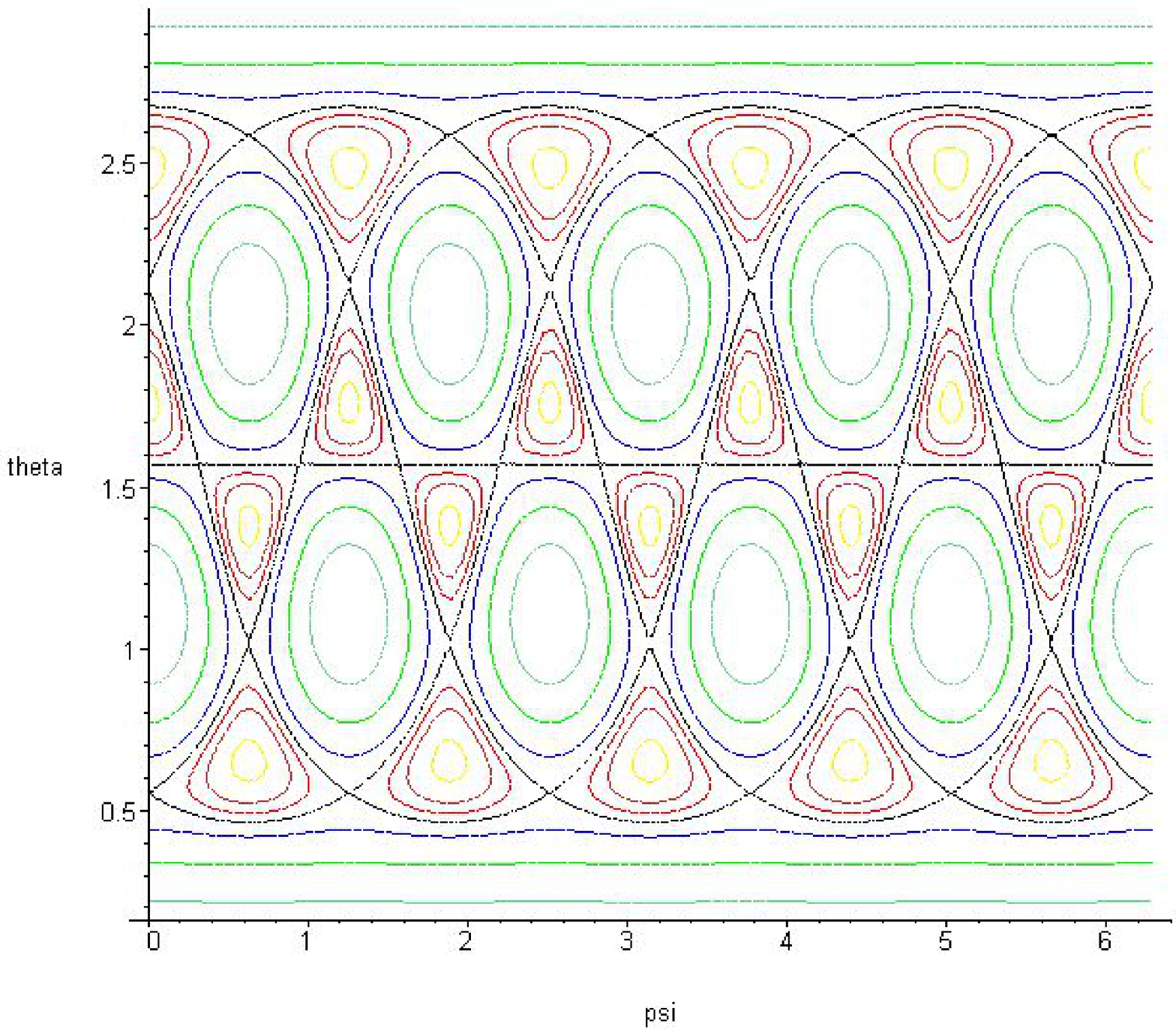}
\end{center}
\caption{{\it  For the potentials $V_1$, $V_2$ and $V_3$ with tetrahedral, cubic-octahedral and dodecahedral-icosahedral symmetry respectively, some isopotential lines are drawn on the sphere $\mathbb S^2$. Aquamarine-green-blue denote decreasing negative and magenta-orange-red increasing positive values of the potential. Black lines denote infinite values of the potential, therefore, they determine regions of the sphere where the motion is confined. Regions with identical behaviour of the potential correspond under the symmetry group of the associated polyhedron. The isopotential lines show the simple structure of the potential in each region of the sphere bounded by the black lines.}
\label{Ft0}}
\end{minipage}
\end{figure}

As a possible explanation of their presumable integrability,  we remark that in each case the  sphere is divided by the singularities of the potential into dynamically separated regions, corresponding each other under the suitable Platonic symmetry group, where the  potential appears to be ``regular" enough to assure integrability. The following figures show the lines of $f_i^{-1}=const.$, respectively, on the sphere, where yellow-red denote positive-increasing values while green-blue the negative-decreasing ones. The black lines denote the zeros of the respective $f_i$ (Fig. \ref{Ft0}).

The research about this topic is in progress.   



\newpage


\end{document}